# THE EVALUATION OF A CODE-SWITCHED SEPEDI-ENGLISH AUTOMATIC SPEECH RECOGNITION SYSTEM

Amanda Phaladi[1] and Thipe Modipa[2]

[1]Department of Computer Science, University of Limpopo
Polokwane, 0727, South Africa
amandamahlako@gmail.com

[2]Centre for Artificial Intelligence Research (CAIR), University of Limpopo, Sovenga,
Polokwane, 0727, South Africa
thipe.modipa@ul.ac.za

## ABSTRACT

*Speech technology is a field that encompasses various techniques and tools used to enable machines to interact with speech, such as automatic speech recognition (ASR), spoken dialog systems, and others, allowing a device to capture spoken words through a microphone from a human speaker. End-to-end approaches such as Connectionist Temporal Classification (CTC) and attention-based methods are the most used for the development of ASR systems. However, these techniques were commonly used for research and development for many high-resourced languages with large amounts of speech data for training and evaluation, leaving low-resource languages relatively underdeveloped. While the CTC method has been successfully used for other languages, its effectiveness for the Sepedi language remains uncertain. In this study, we present the evaluation of the Sepedi-English code-switched automatic speech recognition system. This end-to-end system was developed using the Sepedi Prompted Code Switching corpus and the CTC approach. The performance of the system was evaluated using both the NCHLT Sepedi test corpus and the Sepedi Prompted Code Switching corpus. The model produced the lowest WER of 41.9%, however, the model faced challenges in recognizing the Sepedi only text.*

## 1. INTRODUCTION

For more than five decades, there has been consistent research activity in the field of Automatic Speech Recognition (ASR). ASR, a pivotal technology, plays an important role in enhancing both human-to-human and human-computer interactions [1]. Throughout its history, ASR has consistently propelled the advancement of numerous machine learning (ML) approaches. This includes widely employed methodologies like hidden Markov models, discriminative learning, structured sequence learning, Bayesian learning, and adaptive learning. Furthermore, ASR frequently serves as a substantial and real-world application for speech technology, thoroughly examining the efficacy of specific techniques and inspiring novel challenges that stem from the intricate sequential and dynamic characteristics inherent to speech [2].

Speech recognition technology, as defined by [3], empowers devices to capture spoken words from human speakers using microphones. The realm of speech recognition has witnessed swift advancement in recent years and this momentum continues to build. This technology extends beyond various domains, presenting an array of potential benefits. It finds application in diverse areas, including virtual assistants, automated chatbots, as well as automated transcription and closed captioning for videos [3]. Speech technology encompasses a range of techniques and tools that facilitate machine interaction with other entities through speech.

In recent times, Connectionist Temporal Classification (CTC) has gained significant acknowledgment for its potential to enhance the performance of ASR systems, even in scenarios with limited resources [4]. Various strategies for integrating CTC into ASR models have been put forward in scholarly works. An example of such a strategy involves the incorporation of CTC into the non-autoregressive transformer model with an enhanced decoder input. This method refines the output of a CTC-based model, thereby enhancing the precision of its results [5].

The CTC approach, introduced in [6], empowers deep learning techniques, including Convolutional Neural Networks (CNNs) and Recurrent Neural Networks (RNNs), to play more substantial roles in the field of ASR. In contrast to traditional models that often-output phonemes or other smaller units, CTC directly produces the final target form without requiring additional processing. This streamlined approach simplifies the development and training of end-to-end models, as discussed in [7]. Consequently, it facilitates the creation of a unified network architecture that effortlessly maps input sequences to corresponding label sequences, thereby promoting the advancement of end-to-end speech recognition.

The CTC approach leverages RNNs for sequence labeling tasks where the alignment between input sequences and target labels remains unknown, as high-lighted in [8]. This innovative CTC objective enables the training of end-to-end systems that directly predict grapheme sequences without the need for frame-level alignments of target labels during training. In addition to its simplicity and efficiency, the incorporation of the intermediate CTC loss acts as a regularization technique during training, enhancing performance with minimal code modifications required, as detailed in [9]. Furthermore, CTC imposes minimal overhead during both the training and inference processes, making it a practical choice for ASR systems.

Modipa et al. highlights that a majority of ASR systems heavily rely on transcribed speech data accumulated over an extended period [10]. This data collection aids in establishing statistical associations between sounds in a specific language and learning the underlying patterns. Moreover, it is explained that the Sepedi language encompasses roughly 32 phonemes, while English comprises around 44 phonemes. A phoneme serves as a fundamental unit of sound [11]. As a result, the limited phonemic inventory in Sepedi classifies it as a low-resourced language. Akin to Sepedi, Nguni languages including isiNdebele, isiXhosa, isiZulu, siSwati, as well as the Sotho-Tswana languages, fall under the category of low-resource languages [12]. These languages suffer from various challenges, including limited research attention, scant resources, inadequate digitization, fewer privileges, reduced prevalence in education, or low population density [13], all of which render them relatively underdeveloped compared to languages like English.

This scarcity of resources presents a significant hurdle in developing accurate ASR systems. Biswas [14] points out that the complexity arising from code-switching poses challenges in model integration due to its intricate nature. This becomes particularly formidable when dealing with under-resourced languages, where constraints on textual and acoustic datasets severely constrain modeling capabilities. Thus, our objective is to construct a code-switched language ASR system utilizing an end-to-end methodology, such as the CTC approach. We will utilize this framework to assess ASR performance in code-switched languages, employing two distinct sets of testing data. By training on a diverse multilingual corpus, our system strives to attain high accuracy in recognizing speech within code-switched contexts. We expect that this approach will outperform existing methods in the field of code-switched language recognition, yielding superior ASR results.

This research represents a significant step forward in ASR technology, specifically addressing the complexities of code-switching in under-resourced languages like Sepedi. By developing an efficient and accurate code-switching ASR system, we will contribute to the advancement of speech recognition technology, which has several potential applications in areas such as language learning, speech-to-text transcription, and voice-controlled devices.

In real-world scenarios, particularly in multilingual societies characterized by the prevalent practice of code-switching. It finds practical utility in the context of technologies such as Siri and Google Assistance, which rely on voice interactions and the English language. Given that Africans often incorporate borrowed words from various languages into their speech, our research has the potential to enhance the adaptability and effectiveness of voice-based systems in these linguistic environments.

This study aims to evaluate the impact of varying the number of filters on the modeling of code-switched speech within Automatic Speech Recognition (ASR) systems. The use of different filter numbers plays a crucial role in shaping the model's ability to capture relevant features, patterns, and nuances present in code-switched speech. The number of filters directly affects the model's capacity to discern complex patterns in speech data. By systematically varying the number of filters, this study seeks to uncover the optimal configuration that results in improved accuracy and generalization for transcribing spoken language, particularly in the context of code-switching. The findings aim to provide insights into the refined relationship between filter numbers and the effectiveness of ASR models in handling code-switched speech scenarios. The paper is outlined as follows: Section 2 provides the background of the CTC approach for the development of ASR systems. Section 3 discusses the approach used to develop the Sepedi-English end-to-end system. Section 4 discusses the results obtained. The paper is concluded with Section 5.

## 2. BACKGROUND

In recent years, there has been substantial research and progress in the realm of speech recognition technology, as highlighted in [15]. These advancements have led to increased accuracy and efficiency in ASR systems, owing to developments in machine learning, natural language processing, and deep learning [16]. In the context of code-switched South African speech with limited resources, [17] described the utilization of a hybrid acoustic and language model training technique to enhance ASR accuracy. Additionally, the importance of the Markov assumption in facilitating rapid and parallelized decoding of ASR systems was elucidated in [9]. This assumption empowers the model to compute the probability of the entire output sequence based on the input sequence in a single forward pass, eliminating the need for explicit alignment between input and output sequences.

It has been shown that ASR models utilizing the CTC approach exhibit impressive performance, especially when fine-tuned from wav2vec models, as discussed in [18]. To further improve CTC-based models, researchers have devised two knowledge transfer techniques: representation learning and joint classification learning. These techniques incorporate contextual knowledge from pre-trained language models into ASR systems. Additionally, Lee et al. [9] presents a straightforward yet highly effective auxiliary loss function for ASR based on the CTC objective.

End-to-end models, like CTC, are the predominant choice in the field of ASR. CTC efficiently handles sequential tasks through dynamic programming and effectively leverages Markov assumptions. An alternative approach to aligning acoustic frames with recognized symbols is through the use of an attention mechanism [19]. According to Emiru et al. [20], attention networks offer a promising alternative to recurrent neural networks in end-to-end ASR, showing strong performance. However, it's worth noting that while attention-based models can achieve state-of-the-art results in ASR, they tend to be more intricate and computationally intensive compared to CTC-based models.

While CTC is a commonly used method for ASR in various languages, the primary focus of research and development has largely concentrated on languages with ample speech data available for training and evaluation, as elucidated in [21]. Consequently, assessing the system's performance and scrutinizing the outcomes can offer valuable insights into the effectiveness of CTC for ASR in low-resource languages, while also pinpointing areas for enhancing the Sepedi ASR algorithm.

There is a clear need for further research and development to enhance CTC's performance in low-resource language settings. Moreover, the collection and annotation of speech datasets are imperative to construct a more extensive training corpus.

The Transformers library, introduced by Wolf in 2020, provides support for Transformer architectures and streamlines the distribution of pretrained models [22]. Vaswani et al. [23] introduced the Transformer model in their pioneering work on neural machine translation, wherein an encoder-decoder structure relies solely on attention mechanisms. The Transformer model employs self-attention, in conjunction with other layers within its encoder module, to compute features for word embeddings, harnessing the capabilities of self-attention effectively, as discussed in [24]. Demonstrated findings highlight the remarkable proficiency of transformer-based models in capturing intricate sequential patterns, as evidenced in the studies by [25] and [26].

Transformers, a deep-learning architecture [23] utilize attention-based mechanisms to process sequences. These models employ self-attention modules, allowing them to integrate information from various elements within a sequence during the update process. In contrast to earlier deep-learning methods, transformers exhibit exceptional proficiency in capturing comprehensive dependencies between input and output sequences via attention mechanisms [27]. This enhanced capability translates to improved performance across a wide spectrum of tasks, spanning from natural language processing to computer vision applications. Consequently, transformers have emerged as the preferred architecture for numerous deep learning applications.

The Listen, Attend, and Spell (LAS) model presented by [28] is a neural network designed to convert spoken speech into written characters. This model effectively learns all the integral parts of a speech recognition system in a unified manner. The system is comprised of two main components: a listener and a speller. The listener functions as a pyramidal recurrent network encoder, capable of processing filter bank spectra as its inputs. On the other hand, the speller operates as an attention-based recurrent network decoder, generating character outputs. Notably, this network generates sequences of characters without relying on any presumptions of independence between them. Chen [29] state that while this approach simplifies the training and decoding processes, it becomes challenging for a unified model to adapt when discrepancies exist between the training and testing data, particularly in cases where this information undergoes dynamic changes.

## 3. METHODOLOGY

### 3.1. Code-Switched Speech Corpus

In this research work, we are using the Sepedi Prompted Code Switching (SPCS) corpus [30] and the NCHLT Sepedi corpus. The SPCS corpus is a collection of spoken language data that has been compiled and annotated for linguistic research purposes. The corpus contains recordings of speech from a diverse group of Sepedi speakers aged between 17 and 27 years old. The dataset utilized for SPCS Sepedi was created by capturing spoken expressions from 8 males and 6 females, resulting in a total of 9183 audio files. The SPCS corpus is partitioned into three subsets: 74,6% is allocated for training, 12,4% for validation, and 13.0% for testing purposes.

Table 1 illustrates the allocation of utterances across three distinct sets: training, validation, and testing.

Table 1. The number of utterances of the SPCS corpus.

| Set | Number of utterances |
|---|---|
| Training | 6854 |
| Validation | 1136 |
| Testing | 1193 |

The NCHLT Sepedi corpus NCHLT Sepedi corpus is primarily composed of speech that is initiated through prompts in the Sepedi language. However, it also incorporates instances of English speech, resulting in a code-switched corpus. For our analysis, we focused exclusively on the testing data within the NCHLT corpus, featuring a total of 8 speakers, evenly divided between 4 females and 4 males. We use the testing data extracted from the NCHLT corpus with the testing data obtained from the SPCS corpus, which comprises four speakers, consisting of two females and two males. The distribution of the test data in Table 2 below.

Table 2. Distribution of test data by gender.

| Test data | SPCS speakers | NCHLT speakers |
|---|---|---|
| Female | 2 | 4 |
| Male | 2 | 4 |

During the testing phase, we utilized the NCHLT dataset alongside the SPCS corpus to evaluate the model's performance. This approach allowed us to comprehensively assess the model's capabilities across a range of diverse datasets. The number of utterances in the test data, categorized by gender, can be observed in Table 3 below.

Table 2. Distribution of test data by gender.

| Test data | Number of SPCS utterances | Number of NCHLT utterances |
|---|---|---|
| Female | 662 | 1382 |
| Male | 531 | 1447 |
| **Total** | **1193** | **2829** |

### 3.2 System Development Process

Through a series of sequential actions, we have developed an ASR system utilizing the CTC approach. This progression encompasses several key stages, namely pre-processing, dataset creation, model construction, training, and evaluation. A concise overview of each of these stages follows.

In the initial step, we initiate the process by installing essential Python libraries and modules that are indispensable for the development of the ASR system. This encompasses the incorporation of deep learning frameworks like TensorFlow and Keras. In the pre-processing phase, we begin by specifying the set of characters considered valid within the transcriptions. The resultant character set is then stored within a variable. We proceed to define two Keras layers tasked with facilitating the mapping between characters and integers. The first layer is responsible for translating characters into integers. It utilizes the provided vocabulary argument, which is configured to encompass the pre-defined characters. Any values encountered in the input that are outside this predefined vocabulary are mapped to an empty string. Conversely, the second layer serves as the counterpart, responsible for mapping integers back to their corresponding characters. It leverages the same vocabulary information to perform this mapping operation.

We have established three distinct datasets for the purposes of training, validation, and testing. The testing dataset comprises samples from two separate corpora. To process each encoded sample within the dataset, we employ the map method. This function undertakes tasks such as reading audio files, computing spectrograms, normalizing data, and mapping transcriptions into sequences of integers. To maintain uniform shapes across batches, we apply batching to the dataset, while simultaneously utilizing the pre-fetch method to overlap the pre-processing and training phases. Similarly, the test dataset is constructed, but it employs the test file names and associated transcriptions. This dataset serves the purpose of evaluating the model's performance on a distinct set of data during the training process.

The model configuration followed the end-to-end approach [19]. It embraced a multi-layer architecture, encompassing three RNN layers with 256 units. RNN units in ASR allows the model to analyze and understand sequential data, making them well-suited for converting spoken language into written text. The convolutional aspect involved two 2D CNN layers equipped with 16 filters. The GRU component was characterized by a tanH activation function and a dropout rate of 0.3. The classification layer employed the softmax activation function with a learning rate of 1e-3. We examined the impact of different filter quantities on the convolutional layers to evaluate their influence on recognition accuracy. Specifically, we employed 16, 32, and 64 filters in the study. The number of filters is a hyperparameter that influences the model's capacity to learn and generalize from acoustic features in audio signals. It is often tuned based on the specific characteristics of the dataset. In terms of decoding strategy, a greedy decoding approach was employed, while evaluation employed the CTCLoss function.

Throughout the training process, the neural network receives batches of speech recordings along with their corresponding transcriptions. The network's objective is to discern the inherent patterns within the data and adjust its parameters to minimize the disparity between predicted and actual transcriptions. This optimization is achieved through back-propagation, wherein the model calculates the gradient of the loss function with respect to its parameters and then employs this gradient to adjust the network's weights. This iterative cycle of data ingestion, loss computation, and parameter updating persists for a predetermined number of epochs, typically until the model reaches convergence or the allotted training time expires. In the context of this study, we utilized 150 epochs for training.

### 3.3 Evaluation

Throughout the evaluation process, a callback function is invoked after each epoch, serving the purpose of computing the Word Error Rate (WER) and showcasing a subset of transcriptions from the validation set. This approach provides us with a means to continually assess the model's performance using data that remained unseen during training, allowing for necessary adjustments to the model or hyperparameters. We further evaluate the model's behavior by subjecting it to data from two distinct corpora.

In the model.fit function, the validation data argument designates the dataset used for evaluation during the training process. To clarify, the WER as shown in Eq. (1) quantifies the accuracy of transcriptions, measuring the disparity between predicted words and the actual reference words [31].

$$WER = \frac{S + D + I}{N} \times 100 \qquad (1)$$

where S is the number of substitutions, *D* is the number of deletions, *I* is the number of insertions, *C* is the number of correct words, *N* is the number of words in the reference *(N = S + D + C)*. To calculate the CTC loss the function is as expressed in Eq. (2):

$$P_{CTC}(Y|X) = \sum_{A \epsilon B^{-1}(Y)} \prod_{t-1}^{T} P(a_t|h_t) \qquad (2)$$

for a given input pair *(X, Y)*, we must consider all potential alignments that could result in the output sequence *Y*. We then sum the probabilities of generating each label at each time step, taking into account the corresponding hidden state of the model, as described in [31].

For every alignment *A*, we compute the conditional probability of producing each label at each time step, denoted as $P(a_t|h_t)$. In this context, the output sequence *Y* is represented as a sequence of labels, such as characters or phonemes, with its length indicated as *T*. The input sequence *X* represents the input to the model. A valid alignment *A* serves as a mapping between the input sequence *X* and the output sequence *Y*, while $\sum B^{-1}(Y)$ refers to the set of all valid alignments that result in the output sequence *Y*. By summing the probabilities of generating each label at each time step for all potential alignments, we can determine the CTC loss function.

## 4. RESULTS AND DISCUSSION

In this study, we examined the performance of the ASR model utilizing the CTC approach while altering the number of filters. Two figures, namely Figure 1 and Figure 2, provide a concise summary of the outcomes from this experimentation.

Figure 1 below illustrates the loss across 100 epochs for both training and validation sets. The validation loss pertains to measurements taken on the validation dataset, whereas the training loss corresponds to measurements on the training dataset. The validation loss is a crucial metric that indicates how well the model generalizes to unseen data. Interestingly, the loss shows a decreasing tendency as the number of steps increases, though with occasional fluctuations indicated by spikes. Using 16 filters results in the lowest validation loss with the value 21.96. This suggests that 16 filters work well to improve generality by avoiding overfitting. The loss rises to 22.24 for 32 filters, and to 22.82 for 64 filters respectively.

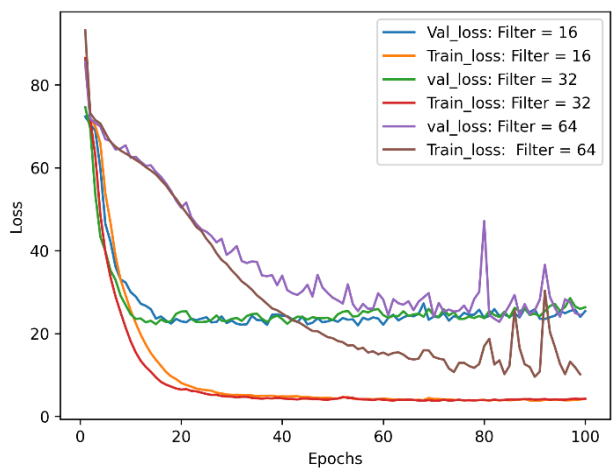

Fig. 1. Validation and training loss by epochs

The findings suggest that, for the given model architecture and dataset, 16 filters strike a good balance between learning from the training data and generalizing to unseen data. Increasing the number of filters to 32 and 64 results in higher validation losses which indicates diminishing returns and potential overfitting. These results emphasize the importance of careful tuning the number of filters, to achieve optimal model performance and prevent overfitting. However, the spikes in the validation loss indicate that the model has difficulty in generalizing to new unseen data at certain epochs, especially with 64 filters, though the smooth training loss plot shows that the model fits the training data well. Additionally, it is noteworthy that for both 16 filters and 32 filter, the validation error exceeds the training error after 10 epochs. For the 64 filter, this deviation appears after 23 periods, which may indicate the beginning of overfitting.

Figure 2 illustrates the Word Error Rate (WER) across different number of filters during the validation phase using the SPCS corpus. The minimum WER is achieved when 16 filters are used, registering at 41.9%. In contrast, when 32 filter are used the WER slightly increased to a WER of 43.48%, and when the 64 filters are used there is a notably elevated WER of 47.47%.

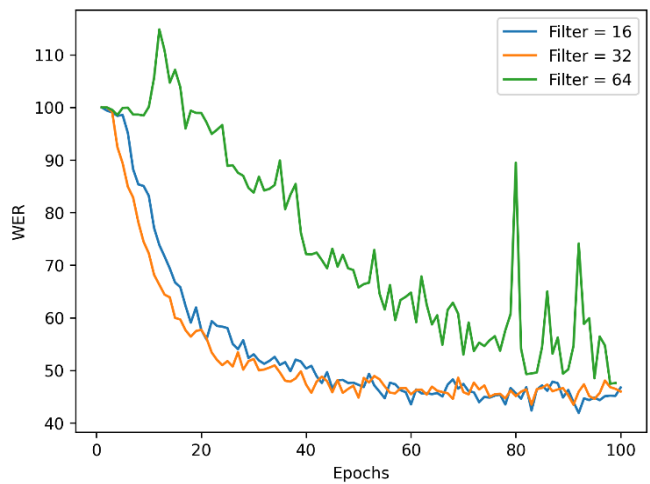

Fig. 2. WER by epochs

The model's performance was best when employing 16 filters, which indicated that for this dataset, a smaller number of filters in the convolutional layers was advantageous. This configuration effectively captured relevant features and patterns in the data. However, as the number of filters increased to 32, a slight rise in WER was observed. This suggests that the model, with increased complexity, started capturing more intricate patterns that did not generalize well to the test set. Furthermore, with 64 filters, the WER notably elevated, indicating that the model had become overly complex, potentially overfitting to the training data and failing to generalize effectively to unseen data. This is because limited training data lead to difficulties in generalizing the unseen data.

When evaluating the SPCS dataset, which serves as the foundation for training, validation, and testing within the study. When subjecting the model to testing using unseen data from the same corpus, the Word Error Rate (WER) exhibits varying results. With 16 filters, the WER is measured at 50.05%, while 32 filters yield a slightly lower WER of 50.24%. Furthermore, employing 64 filters results in a higher WER of 53.51%. When conducting testing with the NCHLT corpus, the smallest WER is 84.59% with 16 filters, followed by 85.06% with 32 filters and 89.89% with 64 filters. These findings underscore the importance of carefully selecting an appropriate number of filters in the model architecture, as the choice significantly influences the model's performance on different corpora. The system lacks training on a varied and representative dataset for the Sepedi NCHLT test corpus, resulting in less favorable outcomes. Hence the NCHLT text corpus encompasses a distinct context compared to the training data, contributing to the observed performance differences. The results make it evident that it is challenging for the model to adapt.

In the study cited as [32], a Word Error Rate (WER) of 55.38% was reported for the testing data involving code-switched Chinese and English. This value exceeds our model's WER by 5.33%, underscoring the superior performance of our model in handling code-switched data, even with a smaller dataset.

Below are the outcomes generated by the model during the testing phase, as demonstrated for both the SPCS and the NCHLT corpus shown in Table 4. The model exhibited a higher degree of accuracy in recognizing the words within the SPCS corpus compared to those within the NCHLT corpus.

Table 4. The target and predicted transcription for SPCS and NCHLT test corpus.

| SPCS test corpus | **Target:** ba re romele di form |
| --- | --- |
| | **Prediction:** ba e romela di form |
| | **Target:** disturba o sa re wa |
| | **Prediction:** disturba o sa re wa |
| **NCHLT test corpus** | **Target:** ke sa le ka go |
| | **Prediction:** ke sa leta go |
| | **Target:** ya ba semaka ge go |
| | **Prediction:** ya ba semaka agego |

The model exhibits proficiency in accurately predicting the initial and concluding words of a given text. Additionally, it demonstrates a high degree of accuracy in forecasting the majority of English sentences. Notably, its performance in predicting longer sentences (consisting of five words) surpasses its accuracy in predicting shorter ones.

## 5. CONCLUSION

This paper presents our efforts in determining the word error rate in speech recognition while varying the number of filters. We conducted experiments using diverse testing datasets sourced from different corpora. The results of our experiments reveal a crucial insight into the selection of the number of filters for optimal model performance. The results demonstrate that the number of filters in a convolutional layer affects the capacity of the network to capture different features in the input data. The extremely high number of filters can have detrimental effects on the model's ability to accurately recognize speech. This emphasizes the significance of achieving a balance in the quantity of filters employed, as excessively high values could result in overfitting, leading to suboptimal performance of the model when exposed to new or varied data.

The Sepedi NCHLT test corpus encapsulated a distinct context in contrast to the training data, posing a challenge for the system to adjust. ASR systems typically exhibit improved performance when the training and testing data closely align in terms of domain and context.

The inclusion of code-switching, where multiple languages are alternated within a conversation, can introduce complexity to ASR systems. If the Sepedi NCHLT test corpus involves complex code-switching patterns than the training data, the system will encounter difficulties accurately recognizing the speech.

Future research directions should include extensive data analysis and experimentation on the Sepedi language, as well as exploration into other under-resourced languages to assess the model's generalizability. Furthermore, the results obtained in this research can act as a fundamental benchmark for upcoming initiatives dedicated to the creation of code-switched Automatic Speech Recognition (ASR) systems for languages that lack adequate resources. These languages face the challenge of limited available data, which is pivotal for training and advancing various language technology systems.

## ACKNOWLEDGEMENTS

I would like to thank my supervisor for guiding me and bringing the best out of me all the time. My gratitude also goes to my family and friends for their support and pushing me during strenuous situations.

**Authors**

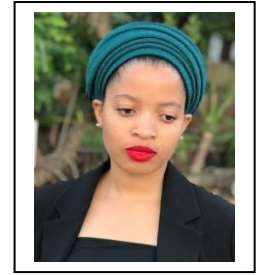

Amanda Phaladi completed her BSc Honours degree at the University of Limpopo and is presently pursuing her MSc degree at the same institution. Her areas of research interest encompass natural language processing, automatic speech recognition, deep learning, and machine learning.

Thipe Modipa earned his PhD in Information Technology from North-West University and currently serves as a member of the Department of Computer Science at the University of Limpopo. His research interests span code-switching for under-resourced languages, text generation, and automatic speech recognition.